\documentclass[twocolumn,amsmath,noshowkeys,showpacs,amssymb]{revtex4-1}

\usepackage{graphicx,color}

\usepackage{amsbsy}
\usepackage{amsthm,mathrsfs,amsopn}

\newtheorem{lemma}{Lemma}
\definecolor{purple}{RGB}{128,0,128}
\definecolor{grey}{RGB}{130,130,130}

\newcommand{\ins}[1]{\textcolor{black}{#1}}
\newcommand{\new}[1]{\textcolor{black}{#1}}

\newcommand{\canc}[1]{}

\begin{document}

\title{Onset of anomalous diffusion from local motion rules}
\author{Sarah de Nigris$^{1,2}$, Timoteo Carletti$^1$, Renaud Lambiotte$^1$}

\affiliation{1. naXys, Namur Center for Complex Systems, UNamur, 5000, Namur, Belgium
\\ 2.  Univ Lyon, Cnrs, ENS de Lyon, Inria, UCB Lyon 1, LIP UMR 5668, 69342, Lyon, FRANCE}

\date{\today}
\begin{abstract} 
Anomalous diffusion processes, in particular superdiffusive ones, are known to be efficient strategies for searching and navigation by animals and also in human mobility. One way to create such regimes are L\'evy flights, where the walkers are allowed to perform jumps,
the ``flights'', that can eventually be very long as their length distribution is asymptotically power-law distributed. 
In our work, we present a model in which walkers are allowed to perform, on a 1D lattice, ``cascades'' of $n$ unitary steps instead
of one jump of a randomly generated length, as in the L\'evy case, where $n$ is drawn from a cascade distribution $p_n$. \new{We} show that this local mechanism may give rise to superdiffusion or normal diffusion when
$p_n$ is distributed as a power law. We also introduce waiting times
that are power-law distributed as well \new{and} \new{therefore the probability distribution scaling is steered by} the two PDF's power-law exponents.
As a perspective, our approach may engender a possible generalization of anomalous diffusion in context where
distances are difficult to define, as in the case of complex networks, and also provide an interesting model for diffusion in  temporal networks.
\end{abstract}


\maketitle

\vspace{0.8cm}

\section{Introduction}
\label{sec:intro}

Diffusion processes, when seen as the continuous limit of a random walk, are well known to display uncanny properties when the associated probability distribution  of length or duration steps for a walker
possesses diverging moments.
Among these unusual diffusion processes, L\'evy flights have been extensively studied on lattices 
and continuous media \cite{metzler2000random,metzler2004restaurant} as they can display \emph{superdiffusion}, so that the variance of the distance covered during the process grows superlinearly 
$\langle x^2(t) \rangle\propto t^\gamma$ 
with $\gamma>1$ at odds with linear diffusion for the Brownian motion \cite{bouchaud1990anomalous,klages2008anomalous}. This enhanced diffusion entails an efficient exploration of the space in which the diffusion process takes place: 
thus both in natural contexts and in artificial ones L\'evy flights have emerged as a strategic choice for such an exploration and 
for search strategies \cite{yang2009cuckoo,sharma2015levy,
boyer2006scale,Haklı2014swarm,
sims2008scaling,deJager2011levy,viswanathan2011physics,mendez2013stochastic,brockmann2006scaling,
gonzalez2008humanmobility,song2010modelling,rhee2011levy,raichlen2014evidence,radicchi2012evolution,radicchi2012rationality,simini2012universal,simini2013human}.
In the case of L\'evy flights, the whole process relies on the divergence of the second moment of the jump probability distribution $P(\ell)$, i.e. the 
probability to perform a jump of length $\ell$. Therefore the walker is allowed to perform 
very long jumps, the flights, which give, as macroscopic effect, the aforementioned superlinear growth of the total displacement
variance $\langle x^2(t) \rangle\propto t^\gamma$ \cite{klafter2011first,shlesinger1993strange}.

On the other hand, if we focus on the temporal properties of the diffusion, we can introduce for the walker a waiting time probability distribution 
$\psi(t)$ determining the probability of jumping after a time $t$ has elapsed since the last move. It is straightforward to see that, assuming its first moment is divergent, a subdiffusive behaviour can emerge 
due to the occurrence of very long waiting times  that slow down the dynamics, i.e. $\langle x^2(t) \rangle\propto t^{\beta}$ with $\beta<1$ \cite{klafter2011first,shlesinger1993strange}.
These two ingredients, the jump length and the waiting time distributions can be blended to create a richer phenomenology 
as it is possible to steer from the subdiffusive regime to the superdiffusive one by tuning the power law distribution exponents of the jump and waiting time probabilities \cite{klafter2011first,magdziarz2007competition}.

\begin{figure}[b]
\includegraphics[width=\columnwidth]{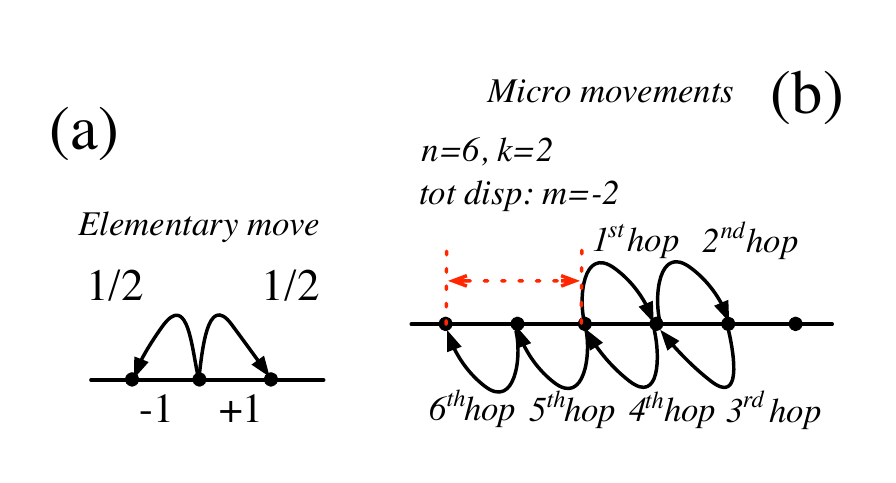}
\caption{The walker is allowed to perform at each step, $n$ elementary moves each one having equal probability $1/2$ to be in the positive or negative direction (Panel a). In panel b) we present a possible avalanche, the walker performs $n=6$ hops, two in the positive direction and $4$ in the negative one, for a total effective displacement of $m=-2$.
\label{fig:hops}}
\end{figure}

In the framework we just described, anomalous diffusion arises from such a choice of the probability distributions for jumps and rest times but it could be unleashed by other properties of the walkers' motion.
In this work, we adopt precisely this perspective: in our model we rely on setting \emph{microscopic} rules for the walker's displacement
so that each ``flight'' is seen as the result as a series of $n$ unitary very small hops, as in Fig.~\ref{fig:hops}.
Anomalous diffusion will therefore stem without the need of an \emph{a priori} knowledge of the jump length \ins{distribution}, as in the canonical L\'evy flight frame,
but it shall be the macroscopic manifestation of such a fragmented and microscopic walk.

The fundamental pivot for the analysis will thus be to relate these microscopic displacements with a macroscopic jump probability distribution 
$P(\ell)$. \ins{For a sake of simplicity}, we investigate this relation on a 1-D chain where we derive an analytical form for the $P(\ell)$ distribution as well as an explicit formula for the displacement variance $\langle x^2(t) \rangle$. \ins{However, we would like to stress that our results could be extended to a more general setting of higher dimensional regular lattices.} Our main result will be that, under suitable 
conditions on the elementary micro-steps distribution $p_n$, the walker 
can indeed exhibit nonlinear diffusion.

The paper is structured as follows: in Sec.~\ref{sec:themodel} we introduce the model and we demonstrate 
that the probability distribution of the jumps $P(\ell)$ can display a divergent second moment. Then, in Sec.~\ref{sec:flights}, we calculate
the probability distribution for the walker that, having in the asymptotic limit a stretched exponential form, leads to a superdiffusive behaviour. 
In Sec.~\ref{sec:simus} we show some numerical simulations \new{to display the L\'evy form of the probability distributions} and we conclude
in Sec.~\ref{sec:conclusion} with some final remarks.
\section{The model}
\label{sec:themodel}
In our model, we consider walkers moving on a $1$-D
lattice able of performing {\em elementary steps} of unitary length, say $+1$ and $-1$, both with
equal probability $1/2$, as shown in Fig.~\ref{fig:hops}.
At each time step, the walker is able to perform $n$ such elementary steps, where $n$ is extracted by some probability distribution function \ins{$p_n$}. In the following we will
assume the latter to follow a power law distribution of exponent $\gamma>1$: 
\begin{equation}
\label{eq:pn}
\ins{p_0\in[0,1]\text{ and }p_n=\frac{C_{\gamma}}{n^{\gamma}}\quad \forall n\geq 1\, ,}
\end{equation}
being $C_{\gamma}=(1-p_0)/\zeta(\gamma)$ a normalising factor and $\zeta(\gamma)$ the Riemann $\zeta$-function. If the probability of not performing any elementary jumps $p_0>0$ then the walker can remain stuck in its current position without doing any elementary steps; on the other hand if $p_0=0$ the walker, at each time step, always performs some elementary jumps, whose possible outcome may eventually be returning to its starting position.  

As we sketched in the Introduction, the pivotal passage for the analysis is to determine the probability $\pi(m)$ to perform a total jump of
length $m$ for some $m\in\mathbb{Z}$ in a time step. Assume the walker performs $n$ elementary steps, then the
probability of making $k\geq 0$ steps in the positive direction, and thus
$n-k$ in the negative one, is given by a binomial process
$\frac{1}{2^n}\binom{n}{k}$, hence the total length will result to be
$m=k-(n-k)=2k-n$. In conclusion we can found: 
\begin{equation}
\label{eq:pm}
\pi(m)=\sum_{n\geq 1}\frac{p_n}{2^n}\binom{n}{\frac{n+m}{2}}+p_0\delta_{m,0}\, ,
\end{equation}
being the last term the probability of performing a total jump of length $m=0$ because the walker did not move at all. The probability to have $m=0$ is composed by this term and an additional one, given by $\sum_{n\geq 1}\frac{p_n}{2^n}\binom{n}{\frac{n}{2}}$, which accounts for the case the walker makes an even number of elementary steps whose total sum is equal to $0$.
Let us observe that, given $m$, not all the values of $n$ do contribute to the
sum: to ensure the positivity of the binomial coefficient, we must
require $n\geq | m|$ and their sum should be an even number, $n+m=2k$,
i.e. they should be both odd or even at the same time. The function $\pi(m)$ is even, as we demonstrate in Appendix~\ref{app:properties}; we can thus restrict ourselves to $m\geq 0$ and rewrite  Eq.~\eqref{eq:pm} for even integers $m=2l$ as follows:
\begin{equation}
\label{eq:pi2l}
\pi(2l)=\sum_{h\geq l}\frac{p_{2h}}{2^{2h}}\binom{2h}{h+l}\quad \forall l\geq 1\quad,
\end{equation}
(note that $h=0$ is not allowed in the sum because it is taken into account thanks to the term $p_0$) and the case $m=0$ reads
\begin{equation}
\quad \pi(0)=\sum_{h\geq 1}\frac{p_{2h}}{2^{2h}}\binom{2h}{h}+p_0.
\end{equation}
For odd integers $m=2l-1$ we obtain
\begin{equation}
\label{eq:pi2lm1}
\pi(2l-1)=\sum_{h\geq l}\frac{p_{2h-1}}{2^{2h-1}}\binom{2h-1}{h+l-1}\quad \forall l\geq 1\,.
\end{equation}
Having computed the probability $\pi(m)$, we now focus on its momenta, in particular the second one as its divergence is known to cause the departure
from normal diffusion \citep{klafter2011first}.
Let $(X_i)_{i\geq 1}$ be independent random variables such that $P(X_i=m)=\pi(m)$, that is $X_i$ is the displacement of the walker at the $i$--th jump, then one can define $T_k=X_1+\dots+X_k$ to be the walker position after $k$ time steps. 
Because of the parity property of $\pi(m)$ one gets $\langle X_k\rangle=0$, an thus $\langle T_k\rangle=0$ for all $k\geq 0$ (see Appedix~\ref{app:properties}).
Using this last remark one can compute the mean square deviation (MSD) as $\mathbb{E}(T_k^2)=\sum_{i\leq k}\mathbb{E}(X^2_i)$ and thus
\begin{align}
\mathbb{E}(X^2_i) &=\sum_m m^2\pi(m)=2\sum_{m\geq 1} m^2\sum_{n\geq m}\frac{p_{n}}{2^n}\binom{n}{\frac{n+m}{2}} \nonumber \\
&=\sum_{n\geq 1}\frac{p_n}{2^{n-1}}\sum_{m=1}^{n}m^2\binom{n}{\frac{n+m}{2}}\,
\label{eq:msd}
\end{align}
where we used the definition of $\pi(m)$ and we rearranged the terms in the sum.
This latter expression acquires a far simpler form (see Lemma~\ref{lem:cn} in Appendix~\ref{app:proof} and the probability distribution Eq.~\eqref{eq:pn}):
\begin{equation}
\mathbb{E}(X^2_i)=\sum_{n\geq 1}np_n=\sum_{n\geq 1}C_{\gamma}\frac{1}{n^{\gamma-1}}\, 
\end{equation}
and thus
\begin{equation}
\mathbb{E}(X^2_i)=
\begin{cases}
C_{\gamma}/C_{\gamma-1}<+\infty &\text{if $\gamma>2$}\\
+\infty &\text{if $1<\gamma\leq 2$}\, .
\end{cases}
\label{eq:divergence}
\end{equation}
In conclusion, if $\gamma>2$ the walker undergoes a linear diffusion process, 
$\mathbb{E}(T_k^2)=kC_{\gamma}/C_{\gamma-1}$. On the other hand, if $1<\gamma\leq 2$ 
we cannot conclude anything using the previous analysis; to overcome this difficulty 
we will consider separately the case $1<\gamma\leq 2$ in the next section.
\new{Before proceeding, we would like to stress that the existence of an interval for the $\gamma$ parameter in which the second moment
diverges is a crucial passage: in our model the walker performs only \emph{local} moves without any a priori knowledge 
of the length it is meant to cover with a jump. This fact paves the way to a generalization 
to contexts in which the space underneath the walker is highly inhomogeneous as we not necessarily require a metric to define the $\pi(m)$.
Therefore, the second moment divergence in our case of study emerges from the interplay of functional form for the $p_n$ and the
topology, in this case a 1-D lattice and, as we show in the next section, this divergence reverberates on the probability distribution itself.}

\section{Discrete time L\'evy flights}
\label{sec:flights}
Although Eq.~\eqref{eq:divergence} has proven the divergence of the MSD when the $p_n\sim 1/n^\gamma$ with $1<\gamma\leq 2$, we do not possess so far any information on how, from a functional perspective, this divergence impacts on the probability distribution. Let us define $P_k(d)$ the probability for the walker to be at distance $d$ from the starting position after exactly $k$ time steps, that is $P_k(d)=P(T_k=d)$. Then using the independence of each jump one can derive the following relation:
\begin{equation}
\label{eq:Pd}
P_{k+1}(d)=\sum_m P_k(d-m)\pi(m)\, ,
\end{equation}
that is the probability to be at distance $d$ at step $k+1$ is given by the probability to be one step before at some position $d-m$ and then make a jump of length $m$. To disentangle this convolution is customary to pass in the Fourier space: 
\begin{equation}
\label{eq:fourier}
\hat{P}_k(\theta)=\sum_dP_k(d)e^{id\theta} \quad\text{and}\quad \lambda(\theta)=\sum_m\pi(m)e^{im\theta}\, .
\end{equation}
Hence using Eqs.~\eqref{eq:Pd} and ~\eqref{eq:fourier}, we obtain
\begin{equation}
\label{eq:Pdfourier}
\hat{P}_{k+1}(\theta)=\hat{P}_{k}(\theta)\lambda(\theta)\, ,
\end{equation}
from which, by iteration, the following expression results
\begin{equation}
\label{eq:hatPkd}
\hat{P}_{k+1}(\theta)=\hat{P}_{0}(\theta)\left(\lambda(\theta)\right)^{k+1}\, ,
\end{equation}
and, applying the inverse Fourier Transform, one can recover $P_k(d)$ from
\begin{equation}
\label{eq:Pkd}
P_k(d)=\frac{1}{2\pi}\int_{-\infty}^{\infty} \left(\lambda(\theta)\right)^ke^{-2\pi id\theta}\, d\theta\,.
\end{equation}
Eq.~\eqref{eq:Pkd} illustrates how, from the behaviour of $\lambda(\theta)$ for $\theta\rightarrow 0$, one can deduce the behaviour of $\hat{P}_{k}(\theta)$ and thus of $P_k(d)$ in the asymptotic limit of large $|d|$. We lever here this standard result to circumvent the divergence in Eq.~\eqref{eq:divergence} and, in order to unveil the divergence rate of the MSD, we shall focus on the behaviour of $\lambda(\theta)$ for small $\theta$ in the following. As we detail in Appendix~\ref{app:lambda}, we are able to explicitly cast it in the form
\begin{equation}
\label{eq:lambdan}
\lambda(\theta)=\sum_{n\geq 1} p_n (\cos\theta)^{n}+p_0\, . 
\end{equation}
Let us define $s=\cos\theta$ and $\mu(s)=\lambda(\cos(\theta))$, then, using the chosen form for $p_n$, we can rewrite Eq.~\eqref{eq:lambdan} as: 
\begin{equation}
\mu(s)=C_{\gamma}\sum_{n\geq 1} \frac{s^{n}}{n^{\gamma}}+p_0\, . 
\end{equation}
To determine the dependence on $s$ in the sum we use the following approximation:
\begin{equation}
\label{eq:sumint}
\sum_{n\geq 1} \frac{s^{n}}{n^{\gamma}} \sim \int_1^{\infty} s^x x^{-\gamma}\, dx\, , 
\end{equation}
for any $s\in(0,1)$- let us remember that we are interested in $\theta \rightarrow 0$ and thus $s\rightarrow 1^-$- we can define $y=-x\log s>0$ and thus change the integration variable form $x$ to $y$:
\begin{align}
\label{eq:sumint2}
&\int_1^{\infty} s^x x^{-\gamma}\, dx = \nonumber \\
&=(-\log s)^{\gamma-1}\int_{-\log s}^{\infty} e^{-y}y^{-\gamma}\, dy=(-\log s)^{\gamma-1} I_{\gamma}(s)\, , 
\end{align}
where $I_{\gamma}(s)$ is defined by the last equality. We note that for $s\rightarrow 1^{-}$ then $I_{\gamma}(s)\rightarrow I_{\gamma}(1)=\Gamma(1-\gamma)$. So, in conclusion we obtain
\begin{equation}
\label{eq:mut0}
\mu(s)-1\sim C_{\gamma}\Gamma(1-\gamma) (-\log s)^{\gamma-1} \quad \text{for}~~s\rightarrow 1^{-},
\end{equation}
where we used the fact that $\mu(1)=1$. Back to $\lambda(\theta)$ we obtain for \ins{$\theta\rightarrow 0$}
\begin{equation}
\label{eq:lambdatheta0}
\lambda(\theta)-1\sim C_{\gamma}\Gamma(1-\gamma) (-\log \cos\theta)^{\gamma-1}\sim \frac{C_{\gamma}\Gamma(1-\gamma)}{2^{\gamma-1}}\theta^{2(\gamma-1)}
\end{equation}
being $\lambda(0)=1$, $\cos\theta\sim 1-\theta^2/2$ and $-\log (1-\theta^2/2)\sim \theta^2/2$.
We thus have, from Eq.~\eqref{eq:hatPkd}, for small $\theta$
\begin{equation}
\ins{\hat{P}_k(\theta)=\lambda^k(\theta)\sim \left( 1-A_\gamma\theta^{2(\gamma-1)} \right)^k \, ,}
\end{equation}
where $A_{\gamma}=-C_{\gamma}\Gamma(1-\gamma)/2^{\gamma-1}>0$.
Therefore, in the limit of large $k$, the above expression tends to the stretched exponential form typical of L\'evy flights characteristic function: 
\begin{equation}
\label{eq:Pktheta2}
\ins{\hat{P_k}}(\theta)\sim e^{-k A_{\gamma}\theta^{2(\gamma-1)}}\, .
\end{equation}
The inverse Fourier Transform of the characteristic function
in Eq.~\eqref{eq:Pktheta2} does not have a straightforward analytical expression and, being 
non-analytic, the evaluation of the MSD using the \ins{standard rule} 
$\left\langle d^2 \right\rangle=\left.\frac{\partial^2}{\partial^2 \theta}\hat{P}_k(\theta)\right|_{\theta=0}$
is impeded because the \ins{latter} expression diverges. It 
is nevertheless possible to exploit the self-similarity property of the distribution \eqref{eq:Pktheta2} 
in order to obtain a scaling relation showing the impact of the local exponent $\gamma$ on the \new{probability distribution}. 
Using \ins{$P_{ak}(a^{1/g}d) = a^{-1/g}P_k(d)$}, we can thus recast the inverse Fourier Transform
\begin{equation}
P_k(d)=\frac{1}{2\pi}\int_{-\infty}^{\infty} e^{-2\pi id\theta-kD_\gamma\theta^{2(\gamma-1)}}\, d\theta\, ,
\end{equation}
\ins{from which we get}
\begin{equation}
\ins{P_k(d)=k^{-1/2(\gamma-1)}\Pi\left(\frac{d}{k^{1/2(\gamma-1)}} \right)\, ,}
\label{eq:p_dt}
\end{equation}
where $\Pi$ is a function of the sole variable $d/k^{1/2(\gamma-1)}$.
\new{We have thus found that the PDF is shaped by the dynamical exponent $\mu=\frac{1}{2(\gamma-1)}$; therefore, as we
anticipated in the previous section, the rule governing the size of the walker's microsteps cascade resonates in 
the overall diffusion process.}
As a closure to the present section we would like to make a remark on the 
probability of \ins{not making any microscopic move} $p_0$. 
{Let us observe that} the walker will always perform
L\'evy flights for any $p_0\in[0, 1)$, being the impact of $p_0$ only on $C_{\gamma}$, more precisely
$C_{\gamma}\rightarrow 0$ when $p_0\rightarrow 1$, but not on the exponent 
$2(\gamma-1)$. Only in the extremal case $p_0=1$ the walk degenerates into an absence of movement.

\section{Continuous time approach}
\label{sec:wt2}
In the previous section we considered a discrete time process in which the steps occurred at a regular pace.
In this section we extend our analysis introducing in our description the waiting time probability distribution, 
which allows the walker to wait \ins{after $n$ micro-steps at the reached position} for a time interval $t$ before hopping again. Thus our process is now
composed by two moves: a waiting time, whose length is weighted by a distribution $\psi(t)$, and a \lq\lq dynamic\rq\rq phase in which $n$ 
elementary steps are instantaneously performed. In our approach, we consider the probability distributions \ins{$\pi(m)$} and $\psi(t)$
as independent and the dynamic phase can be interpreted as the flights in our model since it does not take time, similarly to the classical L\'evy flights.
With these hypotheses, the derivation of the final probability distribution in Fourier-Laplace space $ \hat{P}(\theta,s)$ is straightforward
in the Continuous Time Random Walk (CTRW) frame \cite{klafter2011first}, but we detail here the passages for the sake of completeness.
\begin{figure}[b]
\includegraphics[width=\columnwidth]{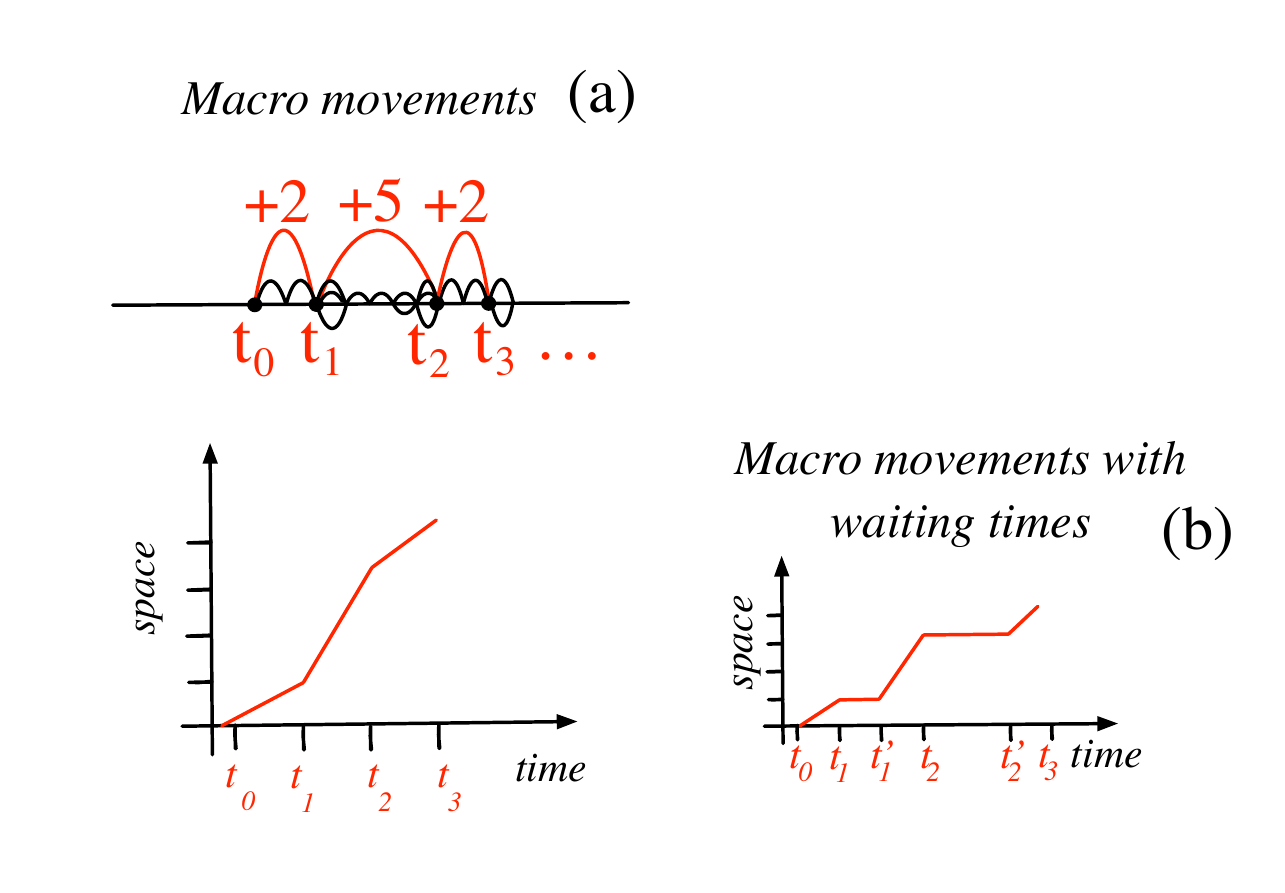}
\caption{In the continuous time frame, at each time $t_i$ the walker can, as before, perform $n$ microscopic moves: in panel a) these macro-movements result in displacements of $m=2$, $m=5$ and $m=2$. If we introduce the waiting times, the displacements are interspersed by waiting intervals: in panel b) the walker stays put from $t_1$ to $t_1'$ and from $t_2$ to $t_2'$.
\label{fig:hops_time}}
\end{figure}
We thus assume that the walker starts at $t=0$ and let $\psi_k(t)$ be the probability distribution function of the occurrence 
of the $k$--th jump at time $t=t_1+\dots+t_k$ where $t_i$ is the waiting time drawn at the $i$--th jump.One clearly has
\begin{equation}
\psi_k(t)=\int_0^t \psi_{k-1}(t')\psi(t-t')\, dt'\, .
\label{eq:psi_k}
\end{equation}
\new{This equation leads, passing in Laplace space (the complete derivation can be found in Ref.~\cite{klafter2011first}), to an expression for  $\chi_k(t)$, which is the probability to make exactly $k$ jumps up to the time $t$.}
\new{In Laplace space, $\tilde{\chi}_k(s)$ reads:}
\begin{equation}
\tilde{\chi}_k(s)=\left(\tilde{\psi}(s)\right)^k\frac{1-\tilde{\psi}(s)}{s}\, .
\end{equation}
Now that the distribution $\chi_k$ accounts for the non-linear relation between steps and time, we
can proceed to include it within the definition of the $P(d,t)$, i.e. the probability for the walker to be at distance $d$ 
from the origin (the initial point at time $t=0$) at time $t$. We observe that this probability is a
generalisation of the previously defined $P_k(d)$: of course, in case all the waiting times are
equal to the duration of the rest period, $\tau$, then $P(d,t)$ reduces to $P_k(d)$ where $k=t/\tau$, as we had in Eq.~\eqref{eq:p_dt}.
Using our starting hypotheses, i.e. that the jumps are costless in time and the waiting time is uncorrelated with the jumps, we can write
\begin{equation}
\label{eq:MEPdt}
P(d,t)=\sum_{k\geq 0} P_k(d)\chi_k(t)\, ,
\end{equation}
meaning that the probability $P(d,t)$ is the probability to be at $d$ after exactly $k$ steps times the probability
to have performed $k$ steps in the time interval $t$.
Using once again the Laplace transform for time, Fourier for space and the result of the previous section we arrive at the classical result
\cite{klafter2011first}:
\begin{align}
\label{eq:MEPdtLap}
\hat{P}(\theta,s)=&\sum_{k\geq 0} \lambda^k(\theta)\tilde{\chi}_k(t)=\sum_{k\geq 0} \lambda^k(\theta)\tilde{\psi}^k(s)\frac{1-\tilde{\psi}(s)}{s}\\\nonumber
&=\frac{1-\tilde{\psi}(s)}{s}\frac{1}{1-\lambda(\theta)\tilde{\psi}(s)}\, .
\end{align}
We thus have that the asymptotic behaviour of $P(d,t)$ shall be governed, in the $d,t\rightarrow\infty$ limit, by the 
moments of the $\lambda(\theta)$ and $\tilde{\psi}(s)$ in the corresponding limit $s,\theta\rightarrow0$ in the Fourier-Laplace space.
Therefore we combine the approximation of $\lambda(\theta)$ in Eq.~\eqref{eq:lambdatheta0} with a waiting time distribution \ins{assumed to have}\canc{displaying
as well} a power-law form $\psi(t)\sim \tau^{\alpha}/t^{1+\alpha}$ with $0<\alpha<1$ for $t\rightarrow\infty$. 
In order to investigate if the waiting time distribution interferes \new{with the PDF's profile}, we focus, for the jump part, on the interesting case where $1<\gamma \leq 2$ as we demonstrated in the previous
section that it leads the second spatial
moment to diverge. {Passing to Fourier-Laplace space, the Laplace transform of $\psi(t)$ reads  $\tilde{\psi}(s)\sim 1-\tau^\alpha s^\alpha$ for small $s$
by virtue of the Tauberian theorem} and, substituting the approximation of $\lambda(\theta)$ and $\tilde{\psi}(s)$ in Eq.~\ \cite{klafter2011first} we obtain
\begin{equation}
\label{eq:MEPdtLapas}
 \hat{P}(\theta,s)\sim\frac{\tau^\alpha s^{\alpha-1}}{\tau^\alpha s^\alpha +A_{\gamma}\theta^{2(\gamma-1)}}\, .
\end{equation}
We then extrapolate the scaling behaviour in the same fashion we derived Eq.~\eqref{eq:p_dt}, where now 
both the exponents $\alpha$ and $\gamma$ intervene in the temporal scaling \cite{fogedby1994langevin}
\begin{equation}
P(d,t)=t^{-\alpha/2(\gamma-1)}\Pi\left(\frac{d}{t^{\alpha/2(\gamma-1)}} \right).
\label{eq:p_dt_alphagamma}
\end{equation}

\section{Numerical simulations}
\label{sec:simus}
\begin{figure}
(a)\includegraphics[width=9.1cm,height=6.6cm]{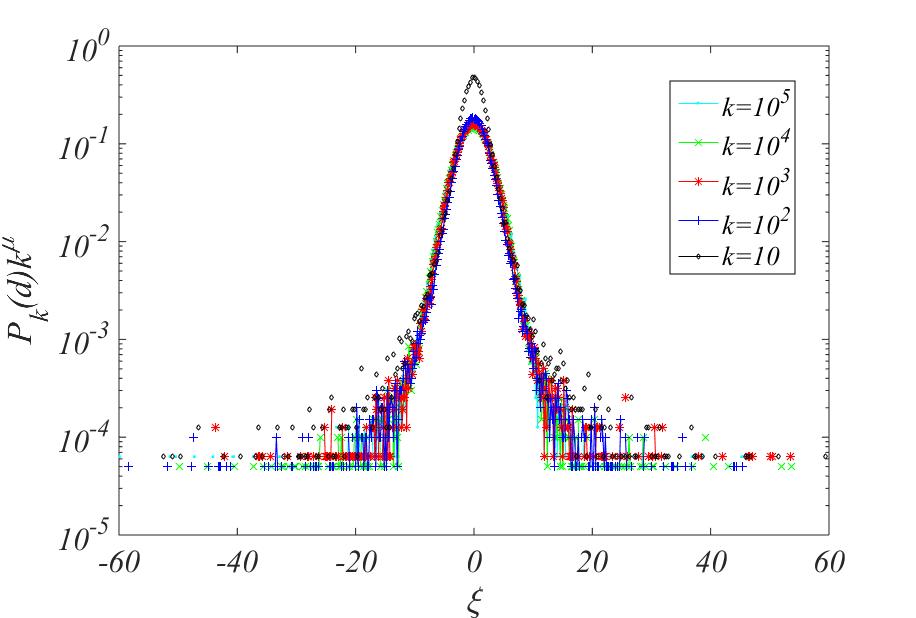}
(b)\includegraphics[width=9cm,height=6.5cm]{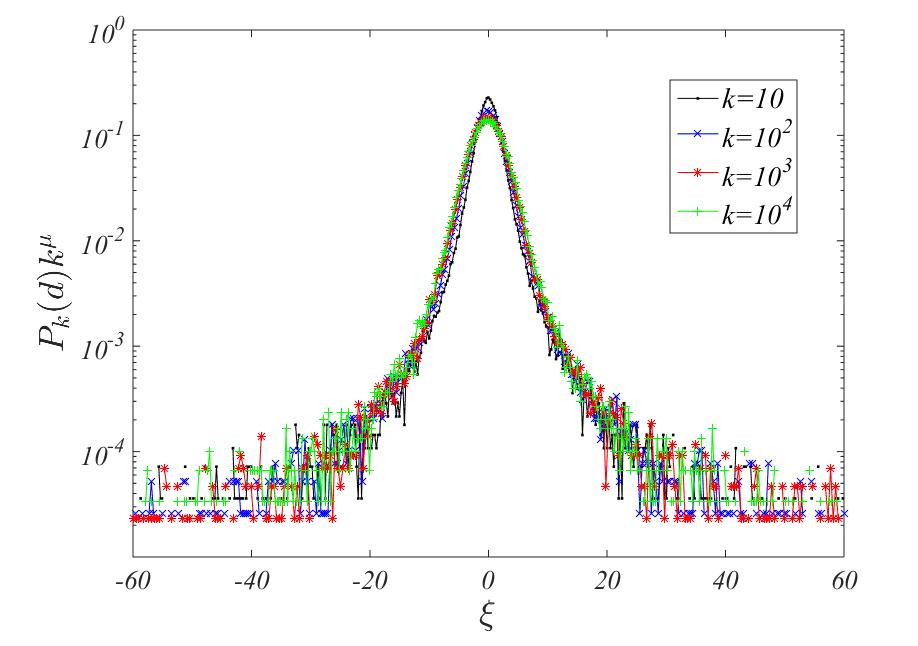}
\caption{\new{Probability distribution $P_k(d)$ in the discrete case for (a) $\gamma=2.1$ and (b) $\gamma=1.9$ 
for $N=10^5$ where $\mu=\frac{1}{2(\gamma-1)}$ and $\xi=\frac{d}{k^{1/2(\gamma-1)}}$. The curves superposition 
 illustrates the self-similarity of the PDF in both cases, but in the (b) the rescaling is reminescent of the local walk properties
 as the variable $\xi$ depends on $\gamma$.
\label{fig:superdiffusive}}}
\end{figure}

\begin{figure}
(a)\includegraphics[trim={11mm 0 0 0},clip,width=9cm,height=7cm]{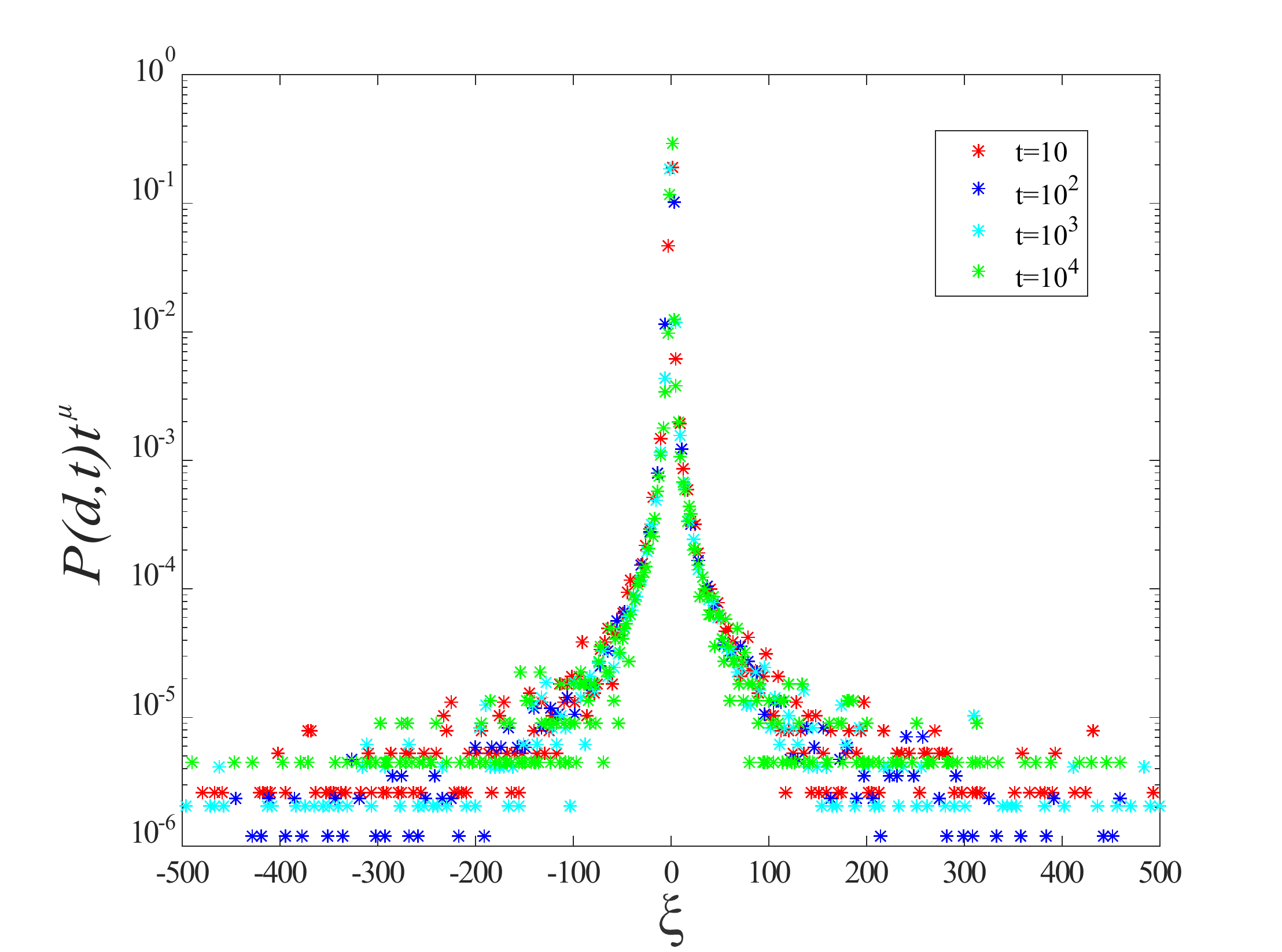}
(b)\includegraphics[trim={7mm 0 0 0},clip,width=9cm,height=7cm]{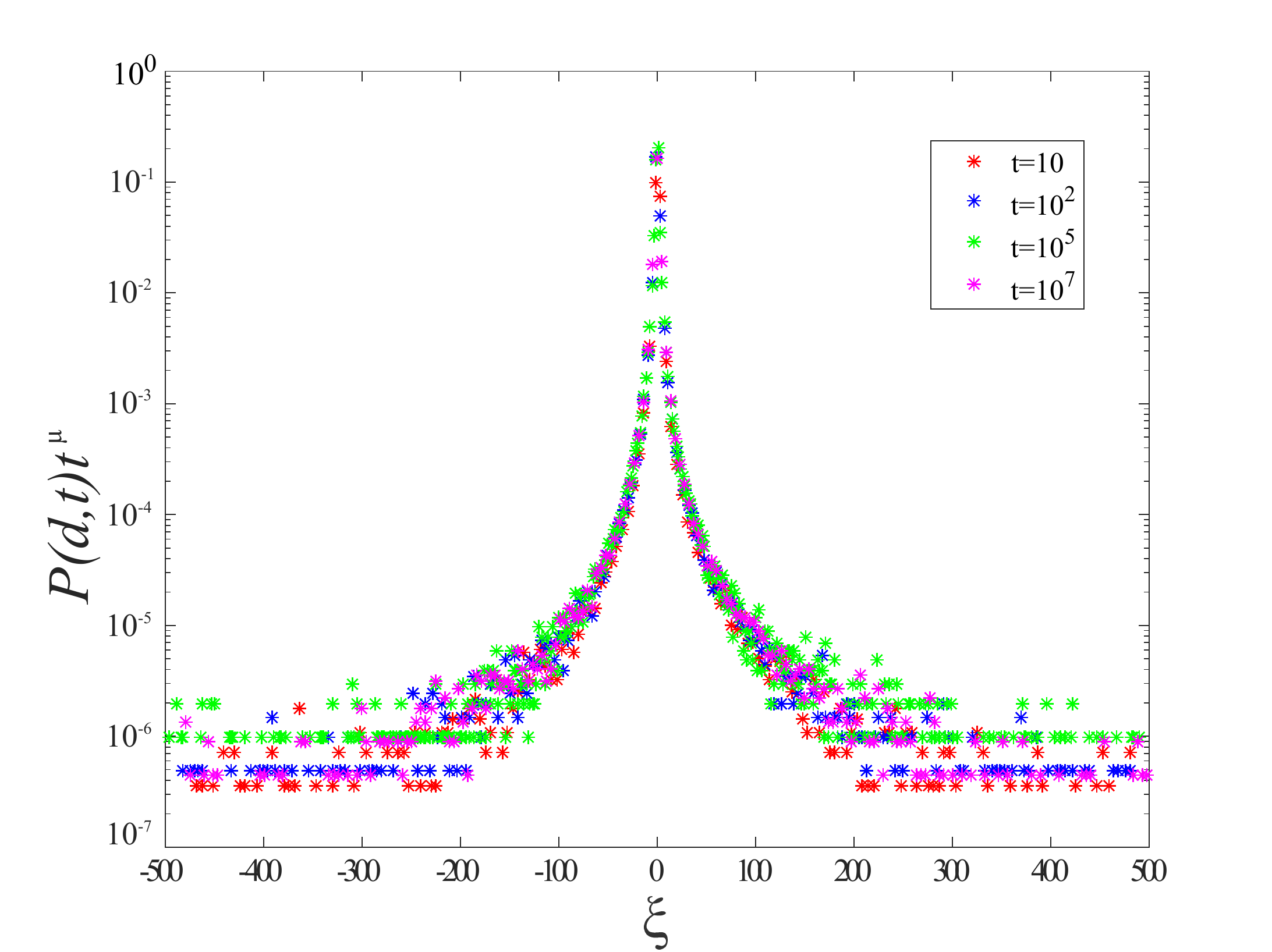}
\caption{\new{Probability distribution $P(t)$ in the continuous time case for (a) $\alpha=0.8$, $\gamma=1.5$ and (b) $\alpha=0.3$, $\gamma=1.7$ for $N=10^5$ where $\mu=\frac{\alpha}{2(\gamma-1)}$ and $\xi=\frac{d}{t^{\alpha/2(\gamma-1)}}$. Again, the self-similarity of the PDF emerges once the curves are rescaled with respect to the $\mu$ exponent.\label{fig:timepdf}}}
\end{figure}

The aim of this section is to present some numerical results to support the theory presented above. We are left now with the numerical 
evaluation of the \new{probability distribution} to confirm the impact brought by the local 
exponents $\alpha$ and $\gamma$ on the overall diffusion process.
\new{As for the discrete case, in Fig.~\ref{fig:superdiffusive}, we show how the $\gamma$ exponent governs the behaviour 
of the probability distribution: indeed as soon as $\gamma>2$, the second moment of the $\pi(d)$ becomes finite and the 
PDF tends to a Gaussian distribution (Fig.~\ref{fig:superdiffusive}a). On the other hand, in the $\gamma\leqslant2$ regime, 
the PDF clearly exibits the fat-tailed L\'evy functional form (Fig.~\ref{fig:superdiffusive}b). In the L\'evy case, 
the variable rescaling that induces the curves collapse in Fig.~\ref{fig:superdiffusive} bears the
mark of the $\gamma$ exponent since we rescale with respect to $\xi=\frac{d}{k^{1/2(\gamma-1)}}$, as obtained in Eq.~\ref{eq:p_dt}. 
For the continuous time regime, both the exponents $\alpha$ and $\gamma$ intervene in the shape of the PDF as shown 
by Eq.~\ref{eq:p_dt_alphagamma}; therefore in Fig.~\ref{fig:timepdf} the superposition of the PDF at 
different times emerges in the same fashion as before once the rescaling is done with respect to the
variable $\xi=\frac{d}{t^{\alpha/2(\gamma-1)}}$.}
As a closing note, it is worth mentioning that other methods exists to tame the numerical instability 
and investigate, albeit indirectly, the theoretical scaling of \new{the $P(d,t)$ moments} such as computing
the fractional moments $\left\langle x^\delta\right\rangle$ with $0<\delta<\mu\leq2$ and $\mu=2(\gamma-1)$ \cite{metzler2000random},
the mean of the displacements $x_i(t)$ logarithm, called the geometric mean $\bar{r_g}$ \cite{lubashevsky2010continuous} and, finally,
computing the probability density averaged within a box with time depending bounds $[L_1t^{1/\mu},L_2 t^{1/\mu}]$ \cite{jespersen1999levy}.

\section{Conclusion}
\label{sec:conclusion}
Concluding, in this work we introduced a random walk model igniting a L\'evy flight type of behaviour and leading to superdiffusion 
on a one dimensional lattice.
The specificity of this model is to impose a {microscopic} condition on the walk, with no need for an a priori 
knowledge of the
topology in order to perform the jumps. In our approach, one jump event corresponds to an ``avalanche'' of $n$ elementary steps,
whose
size $n$ is distributed according to a probability distribution {$p_n$}. We then demonstrated that a power law form
$p_n\sim 1/n^\gamma$ entails the divergence of the second moment of the jumps length distribution \ins{$\pi(m)$} 
when \ins{$1<\gamma\leqslant2$}.
Starting from this divergence, we derived, in Sec.~\ref{sec:flights}, the probability distribution {$\hat{P}_k(\theta)$ in Fourier space}
which has the characteristic stretched exponential form. 
We furthermore introduced the possibility for the walker to stay put on a \canc{node} in Sec.~\ref{sec:wt2}, 
showing \new{how the $\alpha$ exponent of the waiting 
time PDF determines, along with the one of the avalanches $\gamma$, the form of the probability distribution $P(d,t)$.}
Finally, in Sec.\ref{sec:simus} we confirmed through direct numerical simulation the analytical behaviour \new{of the latter, displaying the tails' scaling.}
On a closing note, we would like to stress that the approach itself is independent of the \ins{$p_n$} functional form and that it 
could be generalised to other distributions. The actual meaningful information carried by the \ins{$p_n$}  is the creation
of a divergence in the \ins{jumps second moment computed using the} \ins{$\pi(m)$} distribution. It is worth of
note that 
this divergence stems from the interplay of both the \ins{$p_n$} shape and the $1D$ topology; therefore a careful choice of the former
might be a way to create anomalous diffusion in more general network topologies. 
Widening further our perspective, the walk described in this paper could be used when the underlying space does not possess a proper 
metric and is small-world, as in the case of complex networks \cite{newman2010networks}, such that the probability to perform a walk 
at a certain distance is not univocally defined. In that case, adopting a local perspective for the walker dynamics might prove useful 
to test the notion of anomalous diffusion \cite{riascos2014fractional}. 
Another possible application would the modelling of diffusion on temporal networks \cite{newman2006structure}, especially in the
presence of burstiness \cite{holme2012temporal} and the number of events within a time window can be broadly distributed, possibly 
under the form of trains of events \cite{aoki2016input}.
\acknowledgements
The work of T.C. and R.L. presents research results of the Belgian Network DYSCO (Dynamical Systems, Control, and Optimization), 
funded by the Interuniversity Attraction Poles Programme, initiated by the Belgian State, Science Policy Office.

\bibliographystyle{apsrev4-1}
%

\onecolumngrid
\appendix
\section{Walk properties \label{app:properties}}
\label{app:proof}
\onecolumngrid
The symmetry of the walk reflects in the parity of $\pi(m)$, i.e.  $\pi(-m)=\pi(m)$: 
\begin{align}
\pi(-m) &=\sum_{n\geq |-m|}\frac{p_n}{2^n}\binom{n}{\frac{n-m}{2}} \nonumber \\
&=\sum_{n\geq |m|}\frac{p_n}{2^n}\frac{n!}{\left(\frac{n-m}{2}\right)!\left(\frac{n+m}{2}\right)}=\pi(m)\, .
\end{align}

Therefore, considering the first moment is trivially $\mathbb{E}(X_i)=0$ for all $i\geq 1$:
\begin{align}
\label{eq:EX}
\mathbb{E}(X_i)=\sum_m m\pi(m)=\sum_{m\geq 1} m\pi(m)+\sum_{m\geq 1} (-m)\pi(-m)\new{=0}
\end{align} Hence on average the walker doesn't move from the initial position $\mathbb{E}(T_k)=\sum_{i\leq k}\mathbb{E}(X_i)=0$.
On the other hand, for the MSD, the last equality in Eq.~\eqref{eq:msd} gives
\begin{equation}
\mathbb{E}(X^2_i)=\sum_{n\geq 1}\frac{p_n}{2^{n-1}}\sum_{m=1}^{n}m^2\binom{n}{\frac{n+m}{2}}=\sum_{m=1}^{n}c_np_n\,,
\end{equation}
where
\begin{equation}
c_n\equiv\frac{1}{2^{n-1}}\sum_{m=1}^{n}m^2\binom{n}{\frac{n+m}{2}}\,.
\end{equation}
In order to elucidate its behaviour, we shall use the following Lemma
\begin{lemma}
\label{lem:cn}
Let us define for all $n\geq 1$
\begin{equation}
\label{eq:cn}
c_n=\frac{1}{2^{n-1}}\sum_{m=1}^{n}m^2\binom{n}{\frac{n+m}{2}}\, .
\end{equation}
Then one has
\begin{equation}
\label{eq:cnn}
c_n=n\, .
\end{equation}
\end{lemma}
\proof
Let us consider separately the case $n=2l$ (even) and $n=2l-1$ (odd).

From the definition of $c_n$ and using the parity assumption on $m$ and $n$ we
can rewrite $m=2h$, for some $1\leq h\leq l$, and thus
\begin{equation}
c_{2l}=\frac{1}{2^{2l-1}}\sum_{h=1}^{l}(2h)^2\binom{2l}{h+l}\, .
\end{equation}
The following relations hold to be true
\begin{equation}
\label{eq:binom}
\sum_{k=0}^{p}\binom{p}{k}=2^p \, ,\sum_{k=0}^{p}k\binom{p}{k}=p2^{p-1}
\text{and }\sum_{k=0}^{p}p^2\binom{p}{k}=2^{p-2}(p+p^2)\, .
\end{equation}

Let us develop the definition of $c_{2l}$ to be able to use the previous relations:
\begin{eqnarray}
c_{2l}&=&\frac{8}{2^{2l}}\sum_{j=l+1}^{2l}(j-l)^2\binom{2l}{j}=\frac{8}{2^{2l}}\sum_{j=0}^{2l}(j-l)^2\binom{2l}{j}-\frac{8}{2^{2l}}\sum_{j=0}^{l}(j-l)^2\binom{2l}{j}\\
&=&\frac{8}{2^{2l}}\left[ (2l+4l^2)2^{2l-2}-4l^2
  2^{2l-1}+l^22^{2l}\right]-\frac{8}{2^{2l}}\sum_{j=0}^{l}(j-l)^2\binom{2l}{j}
\\
&=&4l-\frac{8}{2^{2l}}\sum_{j=0}^{l}(j-l)^2\binom{2l}{j}\, ,
\end{eqnarray}
where Eqs.~\eqref{eq:binom} have been used to pass from the first line to the
second one.
Let us rewrite the rightmost term using the change of summing index $j-l=-h$:
\begin{equation}
\frac{8}{2^{2l}}\sum_{j=0}^{l}(j-l)^2\binom{2l}{j}=\frac{8}{2^{2l}}\sum_{h=0}^{l}h^2\binom{2l}{l-h}=\frac{8}{2^{2l}}\sum_{h=1}^{l}h^2\binom{2l}{l+h}=c_{2l}\,
,
\end{equation}
where we used the fact that $\binom{2l}{l-h}=\binom{2l}{l+h}$. In conclusion
we have thus found
\begin{equation}
  c_{2l}=4l-c_{2l}\rightarrow c_{2l}=2l\, .
\end{equation}
The case $n=2l-1$ (odd) can be handled in the same fashion, thus conclud\new{ing} that
\begin{eqnarray}
c_{2l-1}=2l-1\, .
\end{eqnarray}
\endproof
\onecolumngrid
From this equality and the definition of the probability distribution (Eq.~\eqref{eq:pn}), it can be obtained
\begin{equation}
\mathbb{E}(X^2_i)=\sum_{n\geq 1}np_n=\sum_{n\geq 1}C_{\gamma}\frac{1}{n^{\gamma-1}}\, .
\end{equation}
\onecolumngrid

\section{Behaviour of $\lambda(\theta)$ \label{app:lambda}}
\label{app:proof2}
In this appendix we detail the derivation of Eq.~\eqref{eq:lambdan} for the function $\lambda(\theta)$.
Firstly, we observe that using the parity of $\pi(m)$ one can write its Fourier Transform $\lambda(\theta)$ as
\begin{equation}
\label{eq:lambda}
\lambda(\theta)=\pi(0)+2\sum_{m\geq 1}\pi(m)\cos(m\theta)\, ,
\end{equation}
but it is not possible to use the Taylor development $\cos (m\theta)=1-m^2\theta^2/2+\dots$ because in the present case, $1<\gamma\leq 2$, we already know that $\sum_{m\geq 1}\pi(m)m^2$ diverges.
We thus turn to definition of $\pi(m)$ and write
\begin{align}
\label{eq:lambda2}
\lambda(\theta) &=\pi(0)+2\sum_{m\geq 1}\sum_{n\geq m}\frac{p_n}{2^{n}}\binom{n}{\frac{n+m}{2}}\cos(m\theta) \nonumber \\
&=\pi(0)+2\sum_{n\geq 1}p_n\sum_{m=1}^{n}\frac{1}{2^{n}}\binom{n}{\frac{n+m}{2}}\cos(m\theta) \nonumber \\
&=\pi(0)+\sum_{n\geq 1}p_ng_n(\theta)\, ,
\end{align}
where $g_n(\theta)$ is defined using the last equality. For this function $g_n(\theta)$ holds the following lemma
\begin{lemma}
\label{lem:gn}
Let $g_n(\theta)$ be defined by Eq.~\eqref{eq:lambda2}, then $\forall l\geq 1$.
\begin{align}
\label{eq:gn}
&g_{2l}(\theta)=(\cos\theta)^{2l}-\frac{1}{2^{2l}}\binom{2l}{l}\\
&g_{2l-1}(\theta)=(\cos\theta)^{2l-1}\quad \, .
\end{align}
\end{lemma}
\onecolumngrid

%
\proof
Let us consider once again separately the case $n=2l$ (even) and $n=2l-1$ (odd) for some $l\geq 1$. Then one can rewrite
\begin{equation}
\label{eq:gn2}
g_{2l}(\theta)=\frac{1}{2^{2l-1}}\sum_{h=1}^l\binom{2l}{l+h}\cos(2h\theta)\quad \text{and}\quad g_{2l-1}(\theta)=\frac{1}{2^{2l-2}}\sum_{h=1}^l\binom{2l-1}{l+h-1}\cos((2h-1)\theta)\, .
\end{equation}

Let us rewrite the sum for the even case using the variable $j=l+h$ and the sum for the odd case with the variable $j=l+h-1$. \new{Then one
has, for the even case (the odd case can be treated exactly in the same manner)}:
\begin{equation}
\label{eq:gn3}
g_{2l}(\theta)=\frac{1}{2^{2l-1}}\sum_{j=1+l}^{2l}\binom{2l}{j}\cos(2(j-l)\theta)\,.
\end{equation}
Let add and remove in both sums the number of terms up to $h=0$:
\begin{equation}
\label{eq:gn4}
g_{2l}(\theta)=\frac{1}{2^{2l-1}}\left[\sum_{j=0}^{2l}\binom{2l}{j}\cos(2(j-l)\theta)-\sum_{j=0}^{l}\binom{2l}{j}\cos(2(j-l)\theta)\right] \,,
\end{equation}

Rewriting $\cos x=(e^{i x}+e^{-ix})/2$
\begin{eqnarray}
\label{eq:gn5}
g_{2l}(\theta)=\frac{1}{2^{2l-1}}\left[\sum_{j=0}^{2l}\binom{2l}{j}\frac{e^{-2li\theta}(e^{2i\theta})^j+e^{2li\theta}(e^{-2i\theta})^j}{2}-\sum_{j=0}^{l}\binom{2l}{j}\cos(2(j-l)\theta)\right] \, ,
\end{eqnarray}
and using the definition of binomial coefficient $(1+x)^n=\sum_{k=0}^n\binom{n}{k}x^k$ we get:
\begin{eqnarray}
\label{eq:gn6}
g_{2l}(\theta)=\frac{1}{2^{2l-1}}\left[\frac{e^{-2li\theta}(1+e^{2i\theta})^{2l}+e^{2li\theta}(1+e^{-2i\theta})^{2l}}{2}-\sum_{j=0}^{l}\binom{2l}{j}\cos(2(j-l)\theta)\right]\, ,
\end{eqnarray}
and some manipulations give
\begin{eqnarray}
\label{eq:gn7}
g_{2l}(\theta)=2(\cos\theta)^{2l}-\frac{1}{2^{2l-1}}\sum_{j=0}^{l}\binom{2l}{j}\cos(2(j-l)\theta)\, ,
\end{eqnarray}
replacing in both sums $j=l-h$ (and isolating the term $j=0$ in the sum for
even $n$) we get
\begin{eqnarray}
\label{eq:gn8}
g_{2l}(\theta)=2(\cos\theta)^{2l}-\frac{1}{2^{2l-1}}\binom{2l}{l}-\frac{1}{2^{2l-1}}\sum_{h=1}^{l}\binom{2l}{l-h}\cos(-2h\theta)\, ,
\end{eqnarray}
and using the defintion of $g_{2l}$ and $g_{2l-1}$ we obtain:
\begin{eqnarray}
\label{eq:gn9}
g_{2l}(\theta)=2(\cos\theta)^{2l}-\frac{1}{2^{2l-1}}\binom{2l}{l}-g_{2l}(\theta)\, ,
\end{eqnarray}
that is
\begin{eqnarray}
\label{eq:gn10}
g_{2l}(\theta)=(\cos\theta)^{2l}-\frac{1}{2^{2l}}\binom{2l}{l}\, .
\end{eqnarray}
\endproof

Using the previous result we can explicitly rewrite $\lambda(\theta)$ as
\begin{align}
\label{eq:lambdan_app}
\lambda(\theta)&=\sum_{n\geq 1} p_n (\cos\theta)^{n}-\sum_{l\geq
  1}\frac{p_{2l}}{2^{2l}}\binom{2l}{l}+\pi(0)\nonumber\\
  &\equiv \sum_{n\geq 1} p_n (\cos\theta)^{n}+p_0\, ,  
\end{align}
and obtain Eq.~\eqref{eq:lambdan}.
\end{document}